\documentclass{osa-article}
\usepackage{graphicx}
\usepackage{siunitx}

\journal{ome}

\begin{document}

\title{Radially polarized light beams from spin-forbidden dark excitons and trions in monolayer WSe$_2$}

\author{Sven Borghardt,\authormark{1,2,*} Jens Sonntag,\authormark{1,3} Jhih-Sian Tu,\authormark{4} Takashi Taniguchi,\authormark{5} Kenji Watanabe,\authormark{5} Bernd Beschoten,\authormark{3} Christoph Stampfer,\authormark{1,3} and Beata Ewa Kardyna\l\authormark{1,2}}

\address{
\authormark{1}Peter Gr\"unberg Institute 9, Forschungszentrum J\"ulich, DE-52425 J\"ulich, Germany\\
\authormark{2}Department of Physics, RWTH Aachen University, DE-52062 Aachen, Germany\\
\authormark{3}JARA-FIT and 2nd Institute of Physics, RWTH Aachen University, DE-52074 Aachen, Germany\\
\authormark{4}Helmholtz Nano Facility, Forschungszentrum J\"{u}lich, DE-52425 J\"{u}lich, Germany\\
\authormark{5}National Institute for Materials Science, Tsukuba, Ibaraki 305-0044, Japan
} 

\email{\authormark{*}s.borghardt@fz-juelich.de}

%%%%%%%%%%%%%%%%%%% abstract %%%%%%%%%%%%%%%%

\begin{abstract}
The rich optical properties of transition metal dichalcogenide monolayers (TMD-MLs) render
these materials promising candidates for the design of new optoelectronic devices. Despite the large 
number of excitonic complexes in TMD-MLs, the main focus has been put on optically
bright neutral excitons. Spin-forbidden dark excitonic complexes have been addressed
for basic science purposes, but not for applications. We report on spin-forbidden dark excitonic complexes in ML WSe$_2$ as an ideal system for the facile generation of radially polarized light beams. Furthermore, the spatially resolved polarization of photoluminescence beams can be exploited for basic research on excitons in two-dimensional materials.
\end{abstract}

%%%%%%%%%%%%%%%%%%%%%%%%%%  body  %%%%%%%%%%%%%%%%%%%%%%%%%%
\section{Introduction}

The polarization of light acts as the foundation for many physical phenomena and enables a myriad of optical devices. For centuries, the study of polarization was limited to light with spatially homogeneous states of polarization (SOPs), such as linear, elliptical, and circular polarization. Since the end of the last century, light with spatially variant SOPs has been put under the spotlight of intense experimental and theoretical studies \cite{Zhan:09}. In contrast to light with homogeneous SOPs, the polarization in light beams with spatially variant SOPs depends on the position within the cross section of the light beam. Examples for light with spatially variant SOPs are radially and azimuthally polarized light. 

The applications of radially polarized light include, among others, optical trapping\cite{Zhan:04}, improved focusing \cite{QUABIS20001, PhysRevLett.91.233901}, plasmon creation \cite{Chen:07, Kano:98}, and kinematic sensing\cite{Berg-Johansen:15}. While some of these applications require coherent light beams, others work as well with incoherent light. Radially polarized light can be generated actively using, for instance, laser intracavity devices \cite{doi:10.1063/1.1654142} or passively by sending a homogeneously polarized light beam through optical components with spatially variant polarization properties (i.e., phase plates) \cite{Zhan:02, ZHAN2002241}. The realization of compact radially polarized light emitting diodes is highly desirable. However, most semiconductor systems emit either unpolarized or homogeneously polarized light.

Transition metal dichalcogenide monolayers (TMD-MLs) are an emerging class of materials with promising optical properties for next-generation optoelectronic devices\cite{Manzeli2017, RevModPhys.90.021001}. Since the first reports on photoluminescence from MoS$_2$ MLs \cite{PhysRevLett.105.136805, Splendiani2010}, the outstanding optical properties of this material class have been confirmed in a multitude of studies. These include, among others, a direct band gap at the K-points of the hexagonal Brillouin zone \cite{PhysRevLett.105.136805, Splendiani2010}, sizeable spin-orbit coupling in the valence and conduction bands \cite{PhysRevB.84.153402, PhysRevLett.114.046802, Zhang2013, PhysRevB.92.245442, Le_2015}, and massive exciton binding energies \cite{Ugeda2014, PhysRevLett.113.076802, PhysRevLett.113.026803}. Furthermore, the lack of inversion symmetry \cite{PhysRevLett.108.196802} in these materials lifts the spin-degeneracy at the K-points and, thus, induces unique selection rules for optical transitions. These rules give rise to phenomena such as valley polarization \cite{cao2012, PhysRevB.86.081301, Zeng2012, Mak2012, Ersfeld2019Jun, 1911.11692} and valley coherence \cite{jones13}.

The tight excitonic binding in TMD-MLs makes these systems a perfect platform for excitonic light generation. In W-based MLs, for which the uppermost valence band and the lowermost conduction band at the K-points show an anti-parallel spin configuration, the number of excitonic complexes is especially high, and the number of identified ones is growing continuously. So far, experiments and theory have revealed the nature of neutral bright excitons, positively and negatively charged bright excitons (or trions) \cite{PhysRevB.96.085302}, neutral and negatively charged biexcitons \cite{You2015, Paur2019, Ye2018, Barbone2018, Li2018}, as well as spin-forbidden dark excitons \cite{PhysRevLett.119.047401, Zhang2017} and trions \cite{PhysRevLett.123.027401, Li2019, PhysRevB.95.235408}. The spin-forbidden dark states, which do not inherently couple to light, become bright intrinsically through a spin-orbit interaction induced band mixing \cite{PhysRevLett.119.047401} or extrinsically through magnetic field induced spin mixing \cite{Zhang2017, Molas_2017} or near-field coupling to surface plasmon-polaritons \cite{Zhou2017}. An important difference between spin-allowed bright and spin-forbidden dark excitonic complexes in WSe$_2$ MLs is given by the orientation of their transition dipole moments. Bright excitonic complexes recombine via an in-plane transition dipole moment \cite{Schuller2013} while spin-forbidden dark excitonic complexes feature an out-of-plane transition dipole moment \cite{PhysRevLett.119.047401}, resulting in different radiation patterns for bright and dark excitonic complexes \cite{PhysRevLett.119.047401}. As a consequence, a high numerical aperture objective is required in order to probe spin-forbidden dark excitonic complexes when the optical axis of the objective is aligned perpendicular to the ML plane.

To the best of our knowledge, no systematic study on the polarization of the dark states' emission in W-based TMD-MLs has been performed hitherto. In this work, we present spatially resolved measurements of the SOP in light beams resulting from the radiative recombination of dark excitons and trions in a gate-tunable WSe$_2$ MLs and the collection through a high numerical aperture objective. Our accurate measurements show that spin-forbidden dark excitons and trions are ideal sources for the facile generation of radially polarized light beams. The work presented here is relevant for both basic research and applications, as it presents an easy method for the identification of excitonic complexes in two-dimensional materials and proposes a new way for the on-chip generation of coherent vector beams in compact devices easily integratable with optical components such as photonic crystal cavities \cite{doi:10.1063/1.4826679, Ye2015} or photonic waveguides \cite{Zhu2016, Chen2017}.

\section{Gate-dependent photoluminescence} 

\begin{figure}
  \begin{center}
    \includegraphics{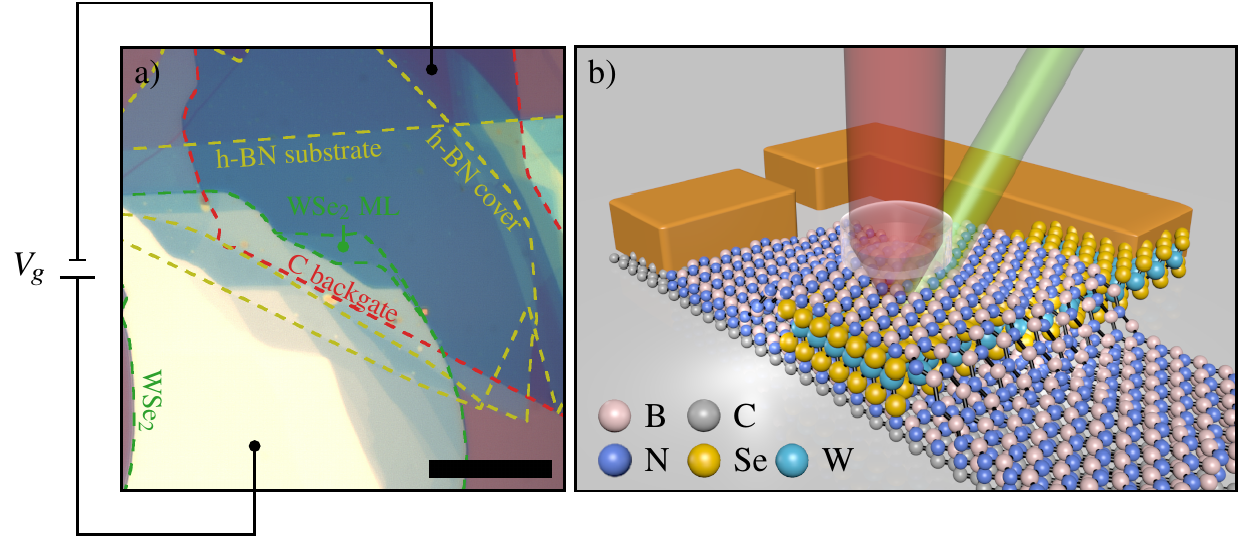}
    \caption{\label{fig:sample} a) Optical micrograph of the gate-tunable WSe$_2$ ML device. Colored dashed lines indicate the edges of the flakes incorporated in the van der Waals heterostructure. The application of a gate voltage was facilitated by providing metal contacts to the few-layer graphene back gate and the few-layer WSe$_2$ flake, as sketched at the exterior of the micrograph. The scale bar is \SI{20}{\micro\metre}. b) Schematic drawing of the $\mu$-photoluminescence experiment performed on the gate-tunable WSe$_2$ ML device. The sample is excited with a continuous wave laser (green), and the photoluminescence signal (red) is collected using an objective with a high numerical aperture. Please note that the exciting beam is focused through the same objective but displaced and tilted in this schematic drawing for illustrative reasons.}
  \end{center}
\end{figure}

We prepared a gate-tunable WSe$_2$ ML device by encapsulating a mechanically exfoliated ML in between two thin flakes of hexagonal boron nitride (h-BN) and placing it onto a few-layer graphene back gate (fig. \ref{fig:sample}) \cite{doi:10.1063/1.4886096}. The encapsulation with atomically flat h-BN flakes has been shown to dramatically improve the optical performance of TMD-MLs, as it mitigates effects from rough substrates and protects the ML from ambiental contamination \cite{PhysRevX.7.021026, Ajayi_2017, Wierzbowski2017}. Furthermore, the lower h-BN flake acts as a high-$k$ dielectric spacer to the few-layer graphene back gate. Metal (Ti/Au) contacts to the few-layer graphene back gate and a thick WSe$_2$ flake attached to the WSe$_2$ ML were defined using electron beam lithography.

Gate-dependent $\mu$-photoluminescence experiments were performed at $T=11$~K. The sample was excited with light from a $1.8$~eV continuous-wave laser, which was focused onto the sample using an aspheric lens with a numerical aperture of $0.72$. The photoluminescence signal from the WSe$_2$ ML was collected through the same objective and analyzed using a silicon CCD at the output of a spectrometer. Voltages between $-0.2$~V and $1.0$~V were applied to the back gate in order to modify the concentration of free charge carriers in the ML. 

\begin{figure}
  \begin{center}
    \includegraphics{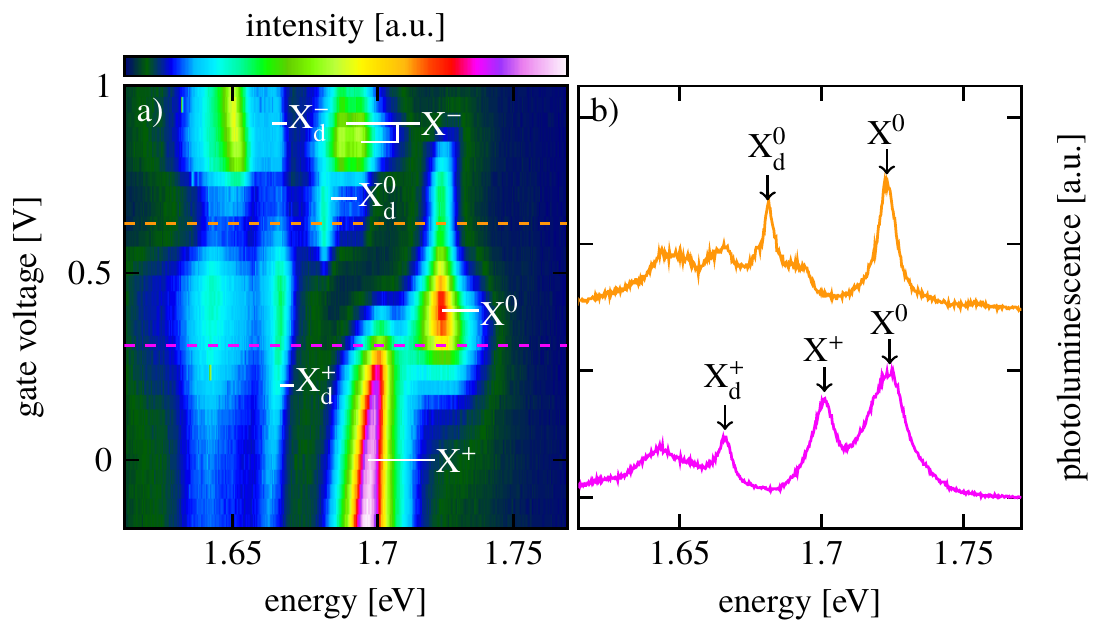}
    \caption{\label{fig:gate_dependence} Gate-dependent photoluminescence from a gate-tunable WSe$_2$ ML. a) shows the photoluminescence spectra in the form of a color map. At the upper (lower) end of the graph, the ML is doped with electrons (holes). The vertical high-intensity lines in the graph represent the different excitonic complexes in the photoluminescence spectra. b) shows two representative photoluminescence spectra for the hole-doped regime ($0.30$~V, magenta line) and the neutral regime ($0.63$~V, orange line). The spectra are normalized and offset for a better comparison.}
  \end{center}
\end{figure}

Figure \ref{fig:gate_dependence} shows the results from the gate-dependent photoluminescence measurements. In figure \ref{fig:gate_dependence} a), multiple excitonic lines can be observed in a photoluminescence map. The neutral bright exciton (X$^0$) recombination is the origin of the highest-energy emission line at $1.72$~eV and is present over a wide range of applied bias voltages. In contrast, the neutral dark exciton (X$^0_\mathrm{d}$, $1.68$~eV), which is probed due to the usage of the high numerical aperture objective, is present only in a small range of voltages close to the neutrality point at approximately $0.63$~V. We define the neutrality point as the gate voltage, at which the collective photoluminescence signal from all charged excitonic states is minimized \cite{1911.01092}. The positive gate-voltage at the neutrality point indicates an inherent hole-doping of the WSe$_2$ ML. In the hole-doped regime, the bright and dark positive trions (X$^+$ and X$^+_\mathrm{d}$) can be observed at $1.70$~eV and $1.66$~eV, respectively. The bright negative trion (X$^-$) and the dark negative trion (X$^-_\mathrm{d}$) generate the highest energy lines in the electron-doped ML. The observed fine structure splitting of the bright negative trion confirms the high quality of our sample \cite{PhysRevB.96.085302}. All measured energies agree well with previous reports on excitonic states in WSe$_2$ MLs encapsulated with h-BN \cite{Ye2018, Barbone2018, PhysRevMaterials.1.054001, 1911.01092}. Multiple broad emission lines appear at low energies but are excluded from the discussion in this work. 

\section{Pinhole scanning}

For the spatially resolved analysis of the SOP in the dark states' photoluminescence signal, the sample was placed in a liquid-helium-cooled cryostat equipped with an objective with a higher numerical aperture of $0.82$. Discrete bias voltages were applied to the back gate in order to probe specifically the $p$-doped and neutral regimes of the WSe$_2$ ML. The sample was excited with right-handed circularly polarized light from a $2.33$~eV fiber-coupled continuous wave laser in order to avoid a linear polarization of the X$^0$ emission caused by valley coherence \cite{jones13}. The linear polarization of the photoluminescence signal was analyzed using a combination of a rotatable half-wave plate and a fixed linear polarizer and then coupled into a multi-mode fiber connected to a spectrometer. Different non-transparent obstacles were introduced in the detection path in order to selectively probe the SOP of specific fractions of the signal beam cross section. 

\begin{figure}[t]
  \begin{center}
    \includegraphics{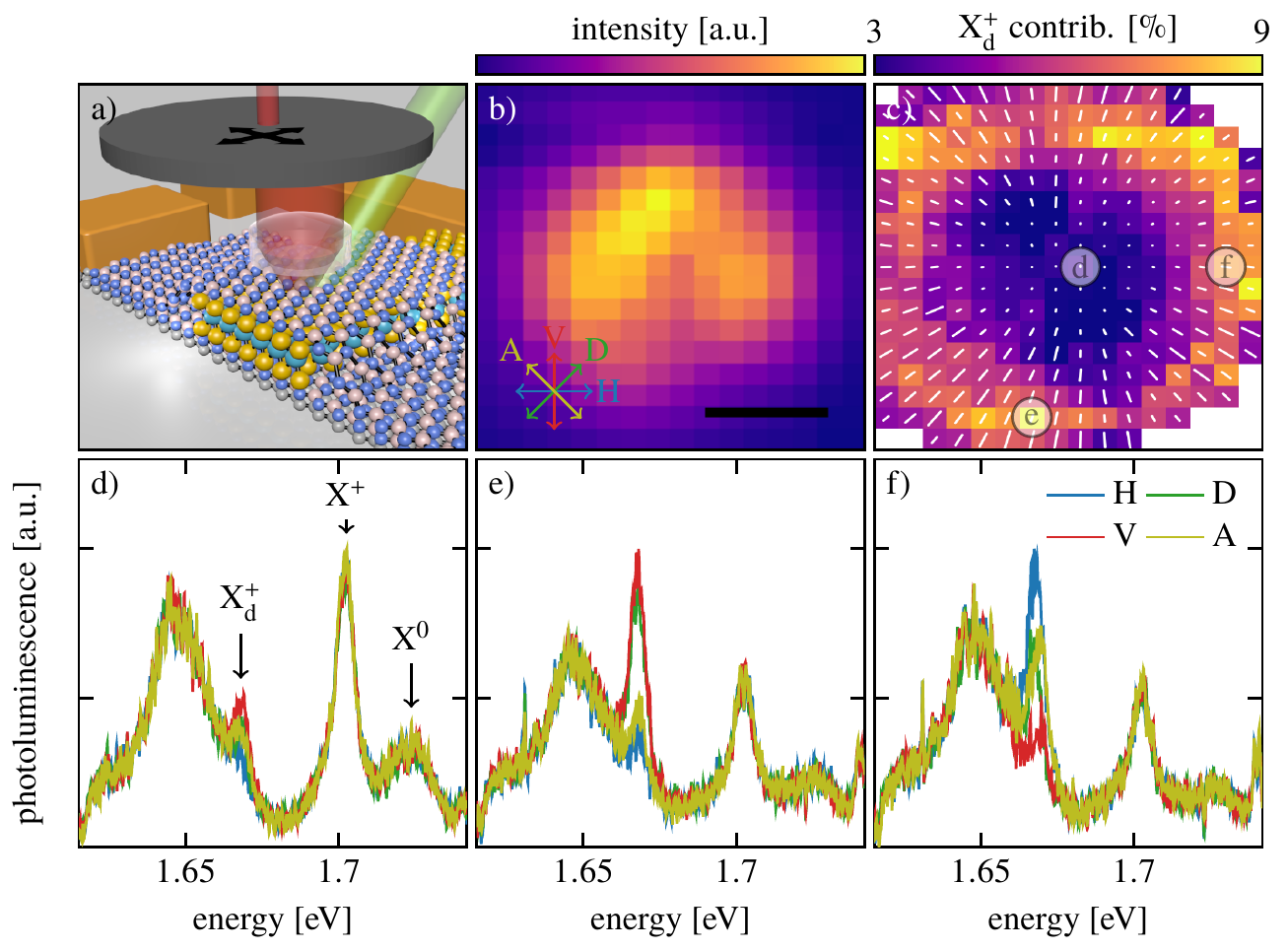}
    \caption{\label{fig:pinhole_trion}Pinhole-scanning experiment for the two-dimensional SOP analysis of the photoluminescence beam from a $p$-doped WSe$_2$ ML. a) shows a schematic drawing of the pinhole, which is scanned across the photoluminescence beam from the WSe$_2$ ML. b) shows the total integrated photoluminescence intensity in the cross section of the signal beam. The scale bar is $1$~mm. The four-axis polarization basis is shown in the lower left corner of the graph; the same color coding of the polarization is used in d)-e). c) shows the contribution of the dark positive trion to the total intensity, as well as its linear polarization. The length of the polarization lines is proportional to the degree of polarization at each position; a polarization line with a length equivalent to the step size in the spatially resolved intensity map (i.e., \SI{200}{\micro\metre}) corresponds to a degree of polarization of $1$. It can be seen that the relative contribution of the X$_\mathrm{d}^+$ emission increases towards the rim of the beam. It is also apparent that the X$_\mathrm{d}^+$ signal beam is radially polarized with a degree of polarization approaching unity at the rim of the beam. d), e), and f) show polarization-resolved spectra at selected positions of the beam cross section. The positions at which the spectra were taken are indicated in c). Note that, for a better comparison, the spectra for different polarizations are normalized to the highest intensity of all spectra shown in a graph. While X$_\mathrm{d}^+$ is hardly polarized and has a relatively low intensity in d), it becomes the dominant emission line in e) and f) with a high degree of polarization.}
  \end{center}
\end{figure}
% changed colors and perspective in a)

A two-dimensionally resolved measurement of the SOP in the photoluminescence beam was performed by introducing a \SI{200}{\micro\metre} pinhole on a biaxial translation stage in the detection path (fig.~\ref{fig:pinhole_trion} a)). The pinhole was scanned across the beam, and spectra in a four-axis polarization basis (horizontal, diagonal, vertical, and anti-diagonal; H, D, V, and A) were acquired for each position of the pinhole by rotating the half-wave plate. Since no linear polarization is expected for the bright excitonic complexes under circularly polarized excitation, the acquired spectra at each position were normalized with the bright exciton and trion intensity in order to cancel out potential instabilities in the measurement.

The two-dimensionally resolved SOP of the photoluminescence beam from a $p$-doped\footnote{The gate voltage applied for this measurement was $V_\mathrm{g}=-1$~V. However, it becomes clear from the photoluminescence spectrum that the charge state of the WSe$_2$ ML is equivalent to $V_\mathrm{g}=0.3$~V in figure \ref{fig:gate_dependence}. The discrepancy results from the transfer to another cryostat and hysteresis effects in the ML.} WSe$_2$ ML is presented in figure \ref{fig:pinhole_trion}. From figure \ref{fig:pinhole_trion} c), it becomes clear that the photoluminescence signal from the dark positive trion shows an intensity distribution which strongly differs from the distribution of the integrated intensity of all excitonic emission lines \ref{fig:pinhole_trion} b). Furthermore, it is evident that the photoluminescence signal beam from the dark positive trion is radially polarized with an increasing degree of polarization as a function of the distance from the beam vortex. We have therefore demonstrated -- for the first time -- that dark positive trions in WSe$_2$ MLs can be used for the generation of radially polarized light beams. Similar results have been obtained for the dark neutral exciton and are presented in the appendix (fig.~\ref{fig:pinhole_exciton}).

The representative polarization-resolved spectra presented in figure \ref{fig:pinhole_trion} d) through f) confirm the conclusions drawn from figure \ref{fig:pinhole_trion} c): While the X$_\mathrm{d}^+$ emission line shows a relatively weak intensity and is almost unpolarized in the center of the beam (fig.~\ref{fig:pinhole_trion} d)), its relative intensity and polarization increase towards the rim of the beam (fig.~\ref{fig:pinhole_trion} e) and f)). Since the pinhole diameter is an order of magnitude smaller than the beam diameter, large fractions of the signal are blocked, leading to a relatively low intensity and, hence, to a relatively low signal-to-noise ratio for the spectra shown in figure \ref{fig:pinhole_trion} d) through f) compared to those presented in figure \ref{fig:gate_dependence} b). Note, however, that a WSe$_2$-based light emitter for radially polarized light would not suffer from this intensity loss.

\begin{figure}
  \begin{center}
    \includegraphics{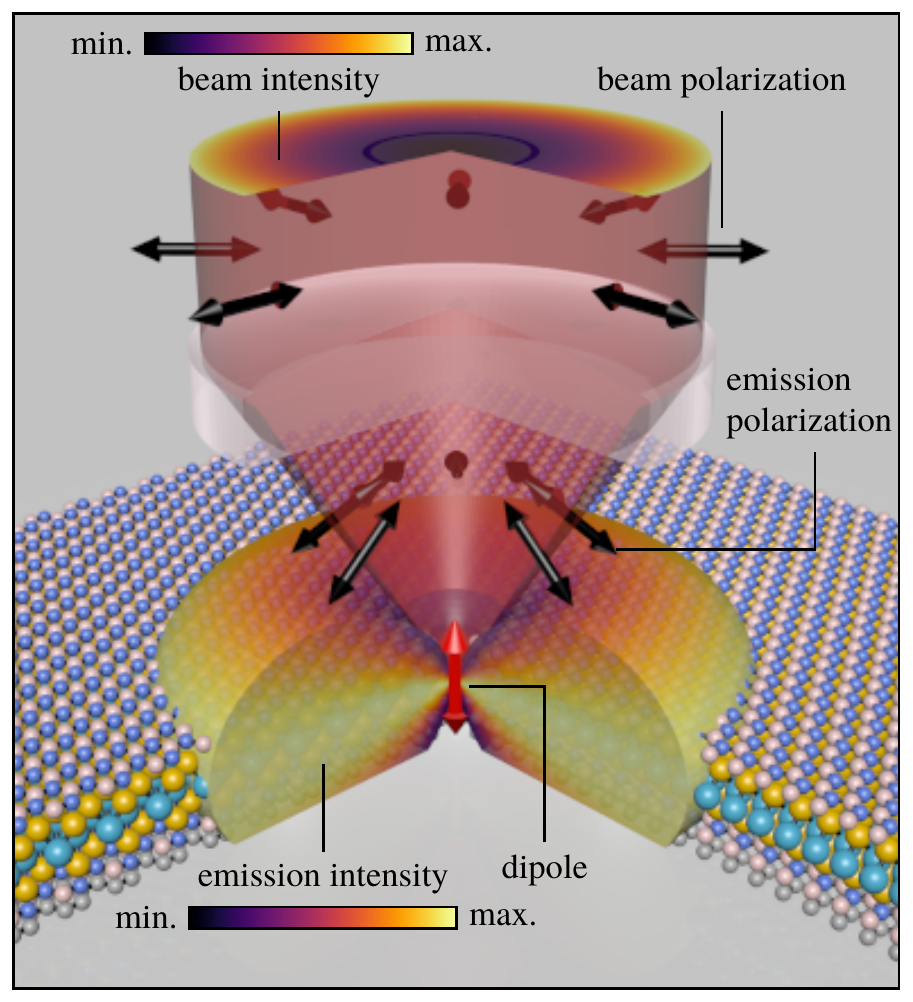}
    \caption{\label{fig:dipole}Collection of light from an out-of-plane dipole using a high numerical aperture objective. The dipole radiation is strongest in the directions perpendicular to the dipole orientation. The polarization of the emitted light is $\pi$-polarized. The high numerical aperture objective collects light emitted at high angles, leading to a high intensity at the rim of the transmitted beam, which is radially polarized.}
  \end{center}
\end{figure}

The spatial distribution of the dark trion's signal beam intensity and SOP are a direct consequence of the electronic band structure of WSe$_2$ MLs and the resulting out-of-plane transition dipole moments for spin-forbidden dark excitonic complexes \cite{PhysRevLett.119.047401}. As depicted in figure \ref{fig:dipole}, the radiation intensity of light emitted from an out-of-plane dipole moment is maximal within the plane of the ML but vanishes in the out-of-plane direction. However, using a high numerical aperture objective (i.e., a large collection angle) light from an out-of-plane dipole emitted into a high angle can be collected. The intensity profile in the signal beam is given by the convolution of the radiation pattern and the collection efficiency of the objective. The light emitted from an out-of-plane dipole is always linearly $\pi$-polarized. At the interfaces formed by the high numerical infinity-corrected objective, the polarization of light is transformed according to Fresnel equations, which produces a radially polarized collimated signal beam.

\section{Knife-edge scanning}

\begin{figure}[t]
  \begin{center}
    \includegraphics{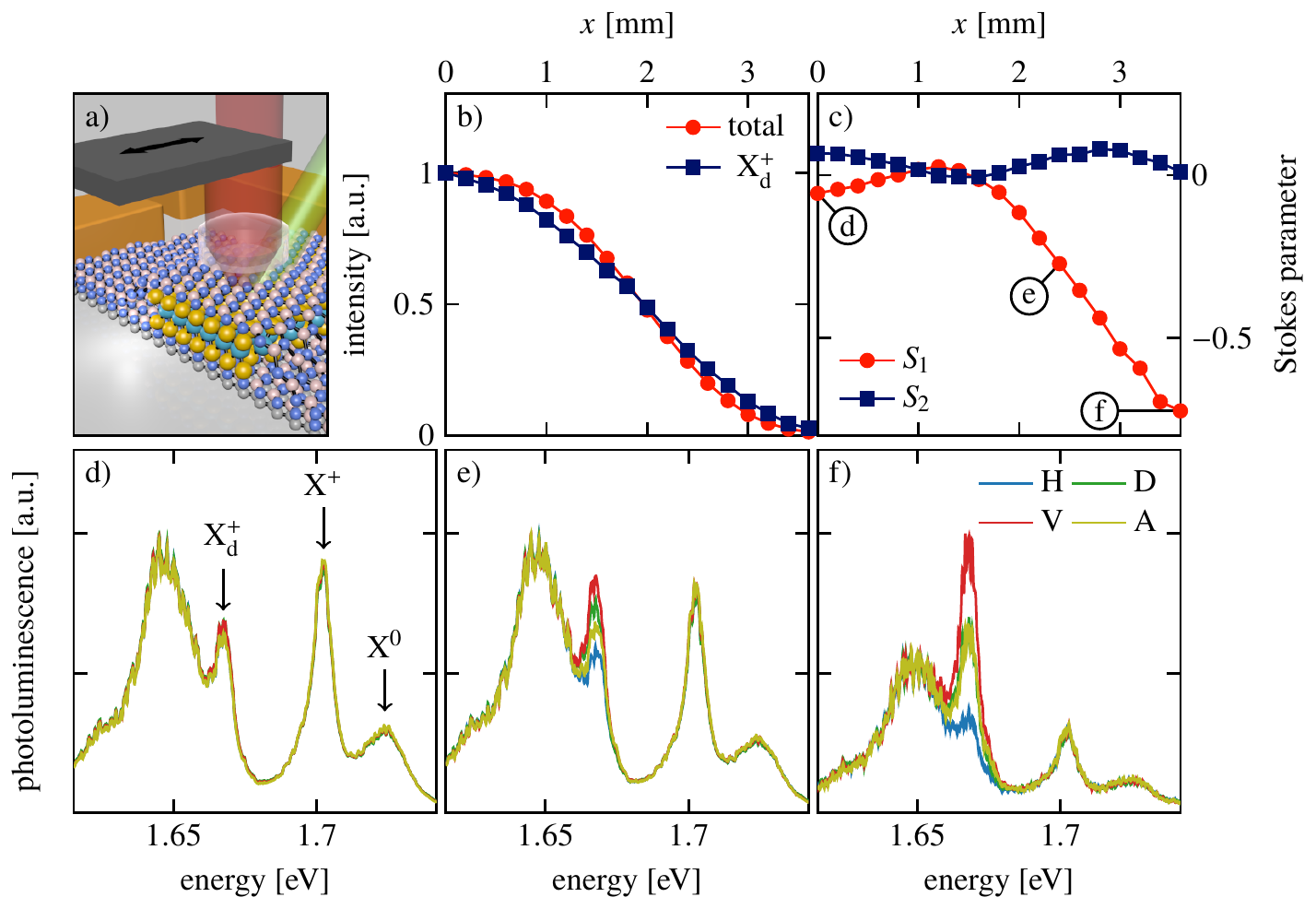}
    \caption{\label{fig:knife_trion}Knife-edge scanning experiment of the photoluminescence beam from a $p$-doped WSe$_2$ ML. a) shows a schematic drawing of the knife-edge, which is scanned across the photoluminescence beam from the WSe$_2$ ML. b) shows the total photoluminescence intensity, as well as the X$_\mathrm{d}^+$ intensity, as a function of the knife-edge position. The intensity profiles have been normalized for a better comparison. c) shows the linear Stokes parameters, $S_1$ and $S_2$, for the X$_\mathrm{d}^+$ signal, as a function of the knife-edge position. d), e) and f) show polarization-resolved photoluminescence spectra for different positions of the knife-edge. The knife-edge positions at which the spectra were taken are indicated in c). The intensity comb observed, for instance, on the emission band at $1.65$~eV results from the operation of the charge coupled device used in the experiment.}
  \end{center}
\end{figure}

The pinhole-scanning experiment provides an accurate measurement of the two-dimensional SOP of a light beam but is time-consuming. For a light beam with rotation symmetry, a knife-edge scanning experiment can provide the same information in a more time-efficient way, although the results are less intuitive. In a knife-edge scanning experiment, a sharp non-transparent obstacle is scanned in one dimension across the signal beam while taking polarization-resolved spectra at each position of the knife-edge (fig.~\ref{fig:knife_trion} a)). In contrast to pinhole-scanning experiments, the time efficiency of knife-edge scanning experiments enables the measurement of the signal beam SOP under varying experimental parameters (e.g., temperature, magnetic field).

Figure \ref{fig:knife_trion} summarizes the results of a knife-edge scanning experiment on the signal beam from a $p$-doped WSe$_2$ ML. In figure \ref{fig:knife_trion} b), the different intensity profiles of the total photoluminescence and the photoluminescence from the dark positive trion in the beam cross section can be seen. Compared to the total photoluminescence intensity, the X$_\mathrm{d}^+$ signal shows a relatively strong decrease when the knife-edge is scanned across the rim of the beam, whereas it shows a relatively small decrease when the knife-edge is passed through the center of the beam. The polarization of the X$_\mathrm{d}^+$ signal is shown in figure \ref{fig:knife_trion} in terms of its Stokes vectors, which are defined as

\begin{equation}
  S_1 = \frac{I_\mathrm{H}-I_\mathrm{V}}{I_\mathrm{H}+I_\mathrm{V}} \qquad S_2 = \frac{I_\mathrm{D}-I_\mathrm{A}}{I_\mathrm{D}+I_\mathrm{A}}
\end{equation}

\noindent with $I_\mathrm{H,D,V,A}$ being the intensity of the subscripted linear polarization component. It is apparent that the knife-edge position has only a small influence on S$_2$, whereas S$_1$ undergoes significant changes with the knife-edge being scanned across the beam. This difference results from the scanning direction of the knife-edge: the knife-edge is scanned in the H-direction and, therefore, equally blocks the A- and D-polarized components of the beam. In contrast, the H- and V-polarized components of the beam are blocked differently. When the knife-edge starts blocking the beam, the H-component on this side of the beam is strongly blocked, whereas the V-component starts being blocked only when the knife-edge approaches the center of the beam. When the knife-edge approaches the outer rim of the beam, the transmitted beam is nearly completely H-polarized. The spectra shown in figures \ref{fig:knife_trion} d) through f) confirm the results from figures \ref{fig:knife_trion} b) and c). When the full beam is transmitted (fig.~\ref{fig:knife_trion} d)), X$_\mathrm{d}^+$ contributes only weakly to the total intensity and is hardly polarized. After passing through the center of the beam, the relative intensity of X$_\mathrm{d}^+$, as well as its degree of polarization, increase continuously (fig.~\ref{fig:knife_trion} e) and f)). Note that, for a perfectly rotation symmetric beam, a completely flat behavior of S$_2$ would be expected. The small oscillations for S$_2$ seen in figure \ref{fig:knife_trion} c) can be attributed to a slightly asymmetric beam resulting, for instance, from inhomogeneities in the sample.

\begin{figure}
  \begin{center}
    \includegraphics{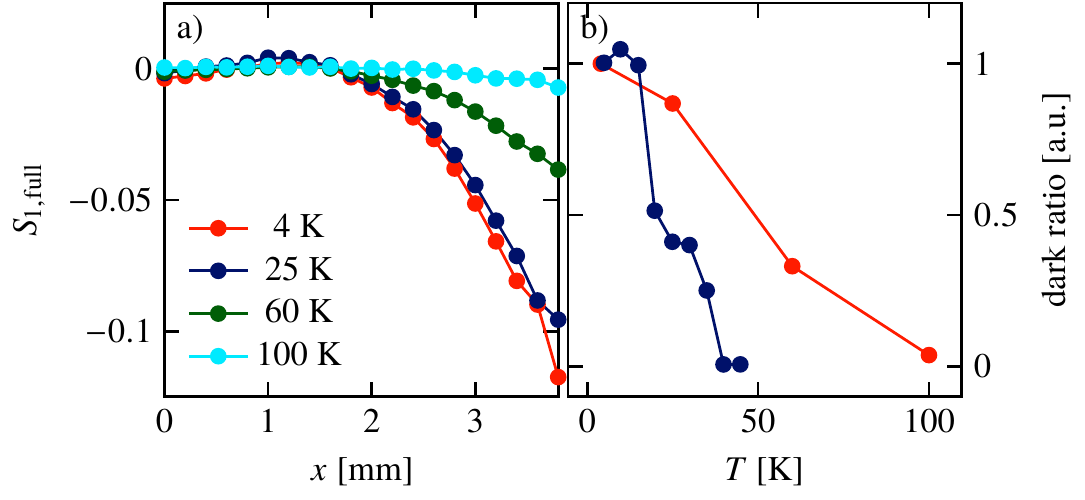}
    \caption{\label{fig:t_dependence}Scanning knife-edge experiments for a $p$-doped WSe$_2$ ML at different temperatures of the WSe$_2$ ML. a) shows the Stokes vector, $S_1$, of the total photoluminescence signal as a function of the knife-edge position for four different temperatures. b) shows the evolution of the contribution from the dark excitonic states to the total photoluminescence signal, as a function of temperature.}
  \end{center}
\end{figure}

The time-efficiency of the knife-edge scanning technique allowed us to use polarization to probe dark excitonic complexes in a systematic way. As an example, we have performed temperature-dependent knife-edge scanning experiments in order to measure the evolution of the dark states' population as a function of temperature. Figure \ref{fig:t_dependence} a) shows the Stokes parameter, $S_1$, of the full integrated signal as a function of the knife-edge position for four different temperatures. It can be seen that the evolution of $S_1$ flattens with increasing temperature. This flattening can be directly attributed to a decrease of the dark states' contribution to the full photoluminescence signal, which can be attributed to a thermally activated depopulation of the dark excitonic complexes. In addition to a pure Boltzmanian redistribution of excitonic complexes, other effects such as, for instance, thermally varying exciton scattering rates can contribute to this depopulation. In figure \ref{fig:t_dependence} b), we show the evolution of the dark states' contribution to the total photoluminescence signal as a function of temperature. The values have been obtained by comparing the $S_1$ parameter for different temperatures at high knife-edge positions ($x>2$~mm). Our measurements show a slower decay of the dark trion intensity with temperature than reported by Liu et al.~\cite{PhysRevLett.123.027401}. This discrepancy is likely to result from differences in the applied experimental parameters (e.g.~charge density or excitation power).

\section{Conclusion and Outlook}

We have provided accurate spatially resolved experimental data on the linear SOP in the photoluminescence signal from gate-tunable WSe$_2$ MLs. Our results prove irrefutably that the signal of spin-forbidden dark excitonic complexes is radially polarized when collected with a high numerical aperture objective with its optical axis perpendicular to the ML plane. The insights provided in our work provide new tools for basic research on low-dimensional excitonic systems and suggest a new application for TMD-MLs.

In terms of basic research, the polarization properties of dark excitons revealed in this work can be applied for the identification of excitonic complexes in two-dimensional materials. For instance, narrow-linewidth quantum emitters in WSe$_2$ MLs have been attributed to spatially confined dark excitons, due to their large $g$-factors \cite{PhysRevLett.123.146401,Lindlau2018}; however, a clear evidence is still missing. The analysis of the SOP of the light from these quantum emitters, e.g., by performing a scanning knife-edge experiment, could shed more light on the character of these states.

In terms of applications, we have shown that WSe$_2$ MLs are an easy-to-use source for radially polarized light beams. However, further engineering of the devices is required for realistic applications. Firstly, since coherence is required for many applications of radially polarized light, cavity engineering is indispensable in order to achieve lasing properties and, thus, coherent emission from dark excitons in WSe$_2$ MLs. Furthermore, electronically driven devices are highly desirable for realistic applications, since they are compact and suppress the signal from the bright excitons, which are populated during optical excitation. Electroluminescence in two-dimensional semiconductors has been be obtained in lateral $p$-$n$-junctions \cite{Ross2014} or vertical van der Waals heterojunctions \cite{Withers2015}. Further engineering of the optical properties of TMD-MLs and, hence, of a potential light source for radially polarized light can be achieved via dielectric engineering \cite{PhysRevMaterials.1.054001, Raja2017} or strain engineering \cite{Castellanos-Gomez2013, He2013, Manzeli2015, Lloyd2016}.

\section{Acknowledgements}

All samples were fabricated in the Helmholtz Nano Facility at Forschungszentrum J\"ulich \cite{hnf}. Growth of hexagonal boron-nitride crystals was supported by the Elemental Strategy Initiative conducted by the MEXT, Japan and the CREST (JPMJCR15F3), JST. This project has received funding from the European Union’s Horizon 2020 research and innovation programme under grant agreement No 785219 (Graphene Flagship).

\newpage
\section*{Appendix}

\begin{figure}[h!]
  \begin{center}
    \includegraphics{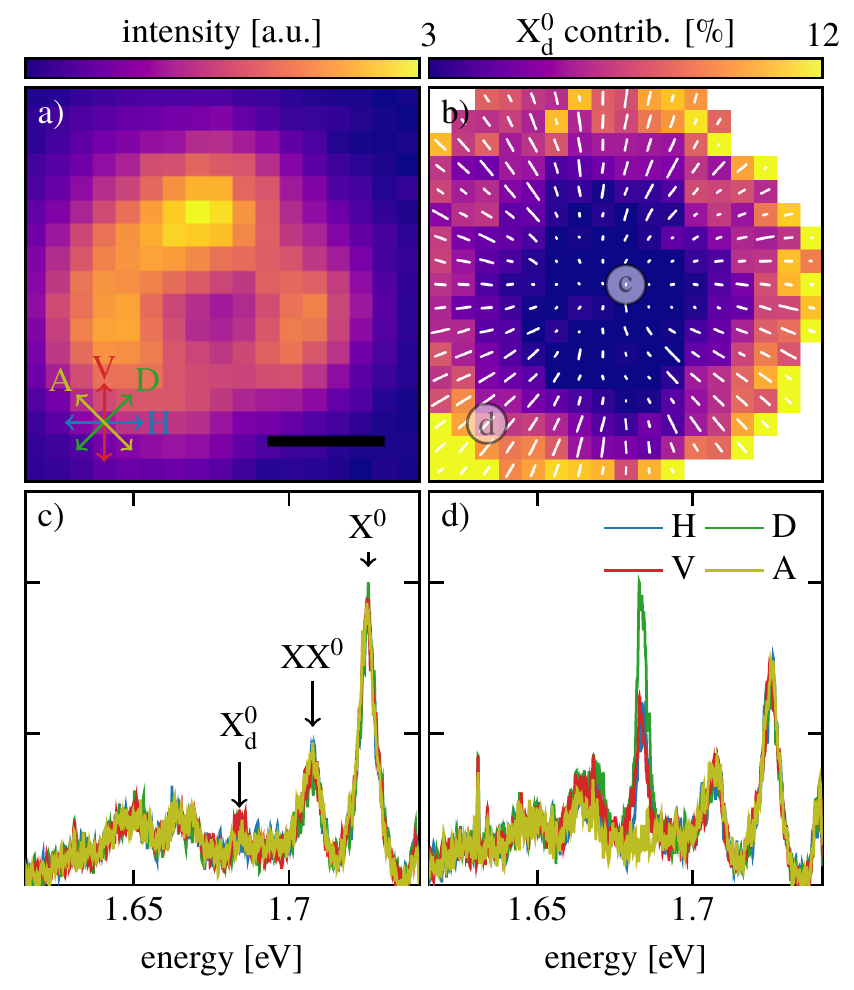}
    \caption{\label{fig:pinhole_exciton}Pinhole-scanning experiment for the two-dimensional SOP analysis of the photoluminescence beam from an undoped WSe$_2$ ML. a) shows the total integrated photoluminescence intensity in the cross section of the signal beam. The scale bar is $1$~mm. The four-axis polarization basis is shown in the lower left corner of the graph. b) shows the contribution of the dark positive exciton to the total intensity, as well as its linear polarization. The length of the polarization lines is proportional to the degree of polarization at each position; a polarization line with a length equivalent to the step size in the spatially resolved intensity map (i.e., \SI{200}{\micro\metre}) corresponds to a degree of polarization of $1$. It can be seen that the relative contribution of the X$_\mathrm{d}^0$ emission increases towards the rim of the beam. It is also apparent that the X$_\mathrm{d}^0$ signal beam is radially polarized with a degree of polarization approaching unity at the rim of the beam. c) and d) show selected polarization resolved spectra at different positions of the beam cross section. The positions at which the spectra were taken are indicated in b). Note that the spectra are normalized for a better comparison. While X$_\mathrm{d}^0$ is hardly polarized and has a relatively low intensity in c), it becomes the dominant emission line in d) with a high degree of polarization. The gate voltage applied for this measurement was $V_\mathrm{g}=-0.34$~V. However, it becomes clear from the photoluminescence spectrum that the charge state of the WSe$_2$ ML is equivalent to $V_\mathrm{g}=0.63$~V in figure \ref{fig:gate_dependence}. The discrepancy results from the transfer to another cryostat and hysteresis effects in the ML.}
  \end{center}
\end{figure}

\clearpage

%%%%%%%%%% If using BibTeX:
\bibliography{references}

\end{document}